\documentclass{edbk} 
\usepackage{edbkps} 
\usepackage{epsfig}     
\let\footnote\savefootnote 
\let\footnoterule\savefootnoterule 
\normallatexbib 
    
\begin{document} 
 
\articletitle{Spectral properties of underdoped~ \\cuprates} 
\author{A. Ram{\v s}ak$^{1,2}$, P. Prelov\v sek$^{1,2}$, and I. Sega$^{1}$} 
\affil{$^1$J. Stefan Institute, SI-1000 Ljubljana, Slovenia\\ 
$^2$Faculty of Mathematics and Physics, University of Ljubljana, 
SI-1000 Ljubljana, Slovenia} 
 
\begin{abstract} 
In the framework of the planar $t$-$J$ model for cuprates we analyze
the development of a pseudo gap in the density of states, which at low
doping starts to emerge for temperatures $T<J$ and persists up to the
optimum doping. The analysis is based on numerical results for
spectral functions obtained with the finite-temperature Lanczos method
for finite two-dimensional clusters.  Numerical results are
additionally compared with the self consistent Born approximation
(SCBA) results for hole-like (photoemission) and electron-like
(inverse photoemission) spectra at $T=0$.  The analysis is suggesting
that the origin of the pseudo gap is in short-range antiferromagnetic
(AFM) spin correlations and strong asymmetry between the hole and
electron spectra in the underdoped regime.

We analyze also the electron momentum distribution function (EMD).
Our analytical results for a single hole in an AFM based on the SCBA
indicate an anomalous momentum dependence of EMD showing "hole
pockets" coexisting with a signature of an emerging large Fermi
surface (FS). The position of the incipient FS and the structure of
the EMD is determined by the momentum of the ground state.  The main
observation is the coexistence of two apparently contradicting FS
scenarios. On the one hand, the $\delta$-function
like contributions at $(\pi/2,\pi/2)$ indicate, that for finite doping a
pocket-like small FS evolves from these points, provided
provided that AFM long range order persists.
On the other hand,
the discontinuity which appears at the  same momentum is more consistent 
with  with infinitesimally short arc (point) of an emerging large
FS.
\end{abstract} 
 
\section{Generalized \lowercase{$t$}-$J$ model} 
  
Spectral properties of underdoped cuprates are of current interest, in
particular results for the electron spectral functions as obtained
with the angle resolved photoemission (ARPES)
\cite{jcc,shen95,norman98}.  A remarkable feature of the ARPES data is
the appearance of a pseudogap already at temperature $T^*$ well above
the superconducting $T_c$.  Other quantities, e.g., the uniform
susceptibility, the Hall constant and the specific heat also show a
pseudogap consistent with energy scale $T^*$
\cite{imada}. At very low temperature $T\ll T^*$ the Fermi surface
and the corresponding electron spectral functions change dramatically with
doping of planar cuprate systems with holes, where a transition from
"small" to "large" FS seems to be consistent with ARPES, but is not
adequately understood from the theoretical point of view.

There have been also several theoretical investigations of this
problem, using the exact diagonalization (ED) of small clusters
\cite{dago,stephan91,eder98,chernyshev98}, string calculations
\cite{eder91}, slave-boson theory \cite{lee96} and the high
temperature expansion \cite{putikka98}.  While a consensus has been
reached about the existence of a large Fermi surface in the
optimum-doped and overdoped materials, in the interpretation of ARPES
experiments on {\it underdoped} cuprates
the issue of the debate is (i) why are
experiments more consistent with the existence of parts of a large FS
-- Fermi arcs or Fermi patches \cite{norman98,furukawa98} -- rather
than with a hole pocket type small FS, predicted by several
theoretical methods based on the existence of AFM long range order in
cuprates, (ii) how does a partial FS eventually evolve with doping
into a large closed one.

The main emphasis of the present study is on the pseudo gap found in
ARPES and also in some exact diagonalization studies
\cite{jprev,jpspec,sigmac}.  We employ the standard $t$-$J$ model to
which we add a nearest neighbor repulsion term $V$,
\begin{eqnarray}
H&=& -t \sum_{<ij>,\sigma}  \bigl(
{ c}_{i,\sigma}^\dagger
{ c}_{j,\sigma} + \mbox{H.c.} \bigr)+ \nonumber \\
& &+\sum_{<ij>}
\bigl[JS_i^z S_j^z+\frac{\gamma}{2}J(S_i^+ S_j^- + S_i^- S_j^+ )
 +  (V -               {J\over 4}) n_{i} n_{j}\bigr].
\label{tjv}
\end{eqnarray}
Here $i,j$ refer to planar sites on a square lattice and $
c_{is}^\dagger$ represent projected fermion operators forbidding
double occupation of sites.  $S^\alpha_i$ are spin operators. For
convenience we treat the anisotropy $\gamma$ as a free parameter, with
$\gamma=0$ in the Ising case, and $\gamma\to1$ in the Heisenberg
model.

Numerical results presented here were obtained with Lanczos exact
diagonalization (ED) technique on small clusters with $N=16 \sim 32$
sites, for temperature in the range $T< J$. The method is simple: we
take into account only the lowest $\approx 100$ Lanczos states
and evaluate the corresponding thermal averages. The method is
compared with a more elaborate finite-temperature Lanczos method
(FTLM)
\cite{jprev} and the agreement in here studied $T<J$
regime is excellent. It should be noted that for low temperatures the
results of the diagonalization of small clusters always have to be
examined with caution, because low energy scale exhibits very strong
finite size effects. 

Our   analytical approach is based on a spinless fermion -- Schwinger
boson representation of the $t$-$J$ Hamiltonian \cite{schmitt88} and
on the SCBA for calculating both the Green's function
\cite{schmitt88,ramsak90,martinez91} and the corresponding wave function
\cite{reiter94,ramsak93}. The method is known to be successful in
determining spectral and other properties of the quasi particles (QP). 
In contrast to other methods the SCBA is expected to correctly describe
the {\it long-wavelength} physics, the latter being determined by the linear 
dispersion of spin waves, whereas the {\it short-wavelength} properties can
be studied with various other methods. Here we compare the SCBA results with
the corresponding ED, as shown further-on.

In the SCBA fermion operators are decoupled into
hole and pseudo spin -- local boson operators: ${
c}_{i,\uparrow}\!=\!h^\dagger_i$, ${
c}_{i,\downarrow}\!=\!h^\dagger_i S^+_i \!\sim\! h^\dagger_i a_i$ and
${ c}_{i,\downarrow}\!=\!h^\dagger_i$, ${
c}_{i,\uparrow}\!=\!h^\dagger_i S^-_i \!\sim\! h^\dagger_i a_i$ for
$i$ belonging to $A$- and $B$-sublattice, respectively.  The effective
Hamiltonian emerges
\begin{equation}
{\tilde H}=
N^{-1/2}\sum_{{{\bf k}}{{\bf q}}} (
M_{{\bf k}{\bf q}}
h_{{{\bf k}}-{{\bf q}} }^\dagger  h_{{{\bf k}} }
\alpha_{{\bf q}}^\dagger+{\rm H.c.} )
+\sum_{\bf q} \omega_{\bf q} \alpha^\dagger_{\bf q} \alpha_{\bf q}
,\label{lsw}
\end{equation}
where $h_{\bf k}^\dagger$ is the creation operator for a (spinless)
hole in a Bloch state. The
AFM boson operator $\alpha^\dagger_{\bf q}$ creates an AFM magnon with
the energy $\omega_{\bf q}$, and $M_{{\bf k}{\bf q}}$ is the
fermion-magnon coupling.

We calculate the Green's function for a hole $G_{\bf k}(\omega)$
within the SCBA \cite{schmitt88,ramsak90,martinez91}.
This approximation amounts to the summation of non-crossing diagrams
to all orders and the corresponding ground state wave function with
momentum ${\bf k}$ and energy $\epsilon_{{\bf k}}$
\cite{reiter94,ramsak93,bala95} is represented as
\begin{eqnarray}
|\Psi_{{\bf k}}\rangle&=&Z^{1/2}_{{\bf k}} \Bigl[
h_{{\bf k}}^\dagger+
...+N^{-n/2}\!\!\!\sum_{{\bf q}_1,...,{\bf q}_n}
\!\!\!
M_{{\bf k}{\bf q}_1}
G_{\bar{\bf k}_1}(\bar\omega_1) \,...\,
M_{\bar{\bf k}_{n-1}{\bf q}_n} \times \nonumber\\
& &\quad\quad \times\;
G_{\bar{\bf k}_n}(\bar\omega_n)
\;h_{\bar{\bf k}_n}^\dagger
\alpha_{{\bf q}_1}^\dagger...\,\alpha_{{\bf q}_n}^\dagger  +\,\,\,
... \,\,\,
\Bigr]
|0\rangle. \label{psi}
\end{eqnarray}
Here $\bar{\bf k}_m={\bf k}\!-\!{\bf q}_1\!-\!...\!-\!{\bf q}_m$,
$\bar\omega_m=\epsilon_{{\bf k}}\!-\!\omega_{{\bf q}_1}\!-\!...\!-\!
\omega_{{\bf q}_m}$ and
$Z_{{\bf k}}$ is  the QP spectral weight.

\section{Pseudo gap in the density of states}
      
We study here the planar density of states (DOS), 
defined as ${{\cal N}}(\omega) = 2/ N
\sum_{{\bf k}} { A}_{\bf k}(\omega-\mu)$, where $A_{\bf k}(\omega)$ is
the electron spectral function \cite{jpspec,jprev}, and $\mu$ denotes the
chemical potential. First we calculate the DOS with the
finite-temperature Lanczos method for clusters of $N=18,
20$ sites doped with one hole, $N_h=1$. Here we denote with ${{\cal
N}^-}(\omega)$ the density of states corresponding 
to adding a hole into the system and thus
to the photoemission experiments, while ${{\cal N}^+}(\omega)$
represents the inverse photoemission (IPES) spectra.
       
\begin{figure}[htb]   
\epsfig{file=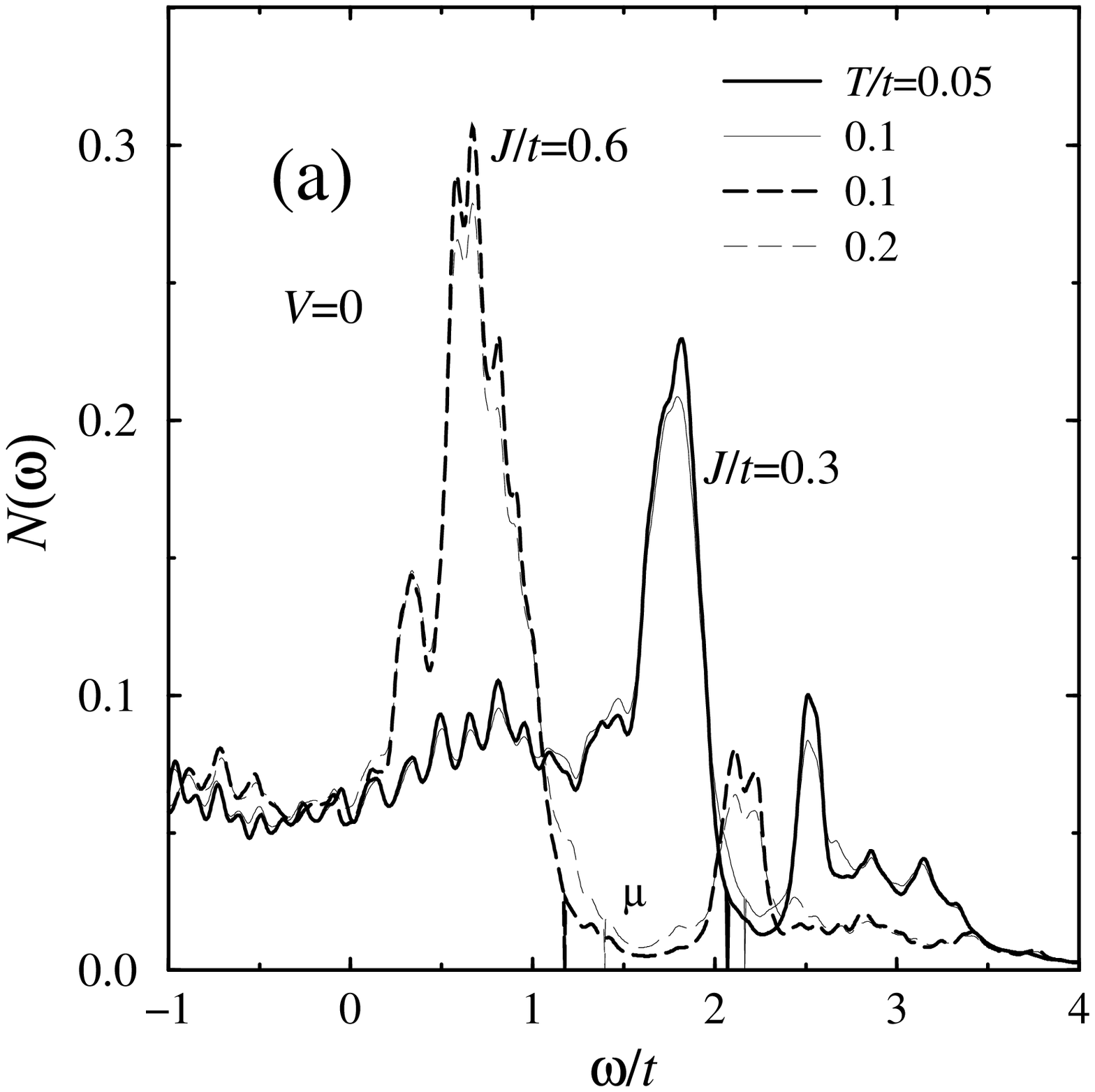,height=60mm,angle=-90}
\epsfig{file=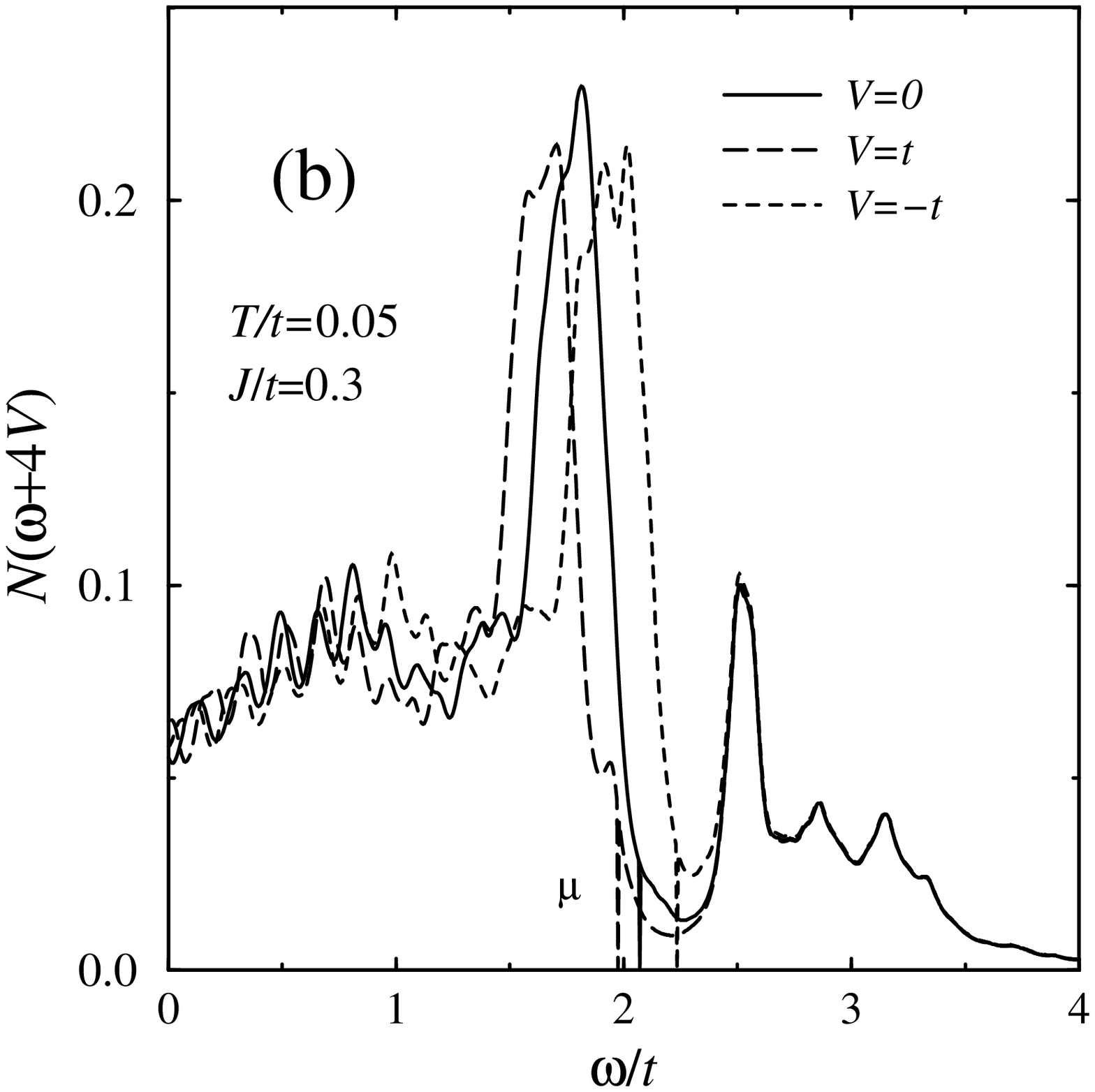,height=60mm,angle=-90}
\caption{${\cal N}(\omega)$ for one hole on $N=18$ sites, presented
for different $J/t$, $V/t$ and $T/t$. (a) $V=0$. (b) $J=0.3t$.
Broadening of peaks is taken $\delta/t=0.04$.}
\label{fig1} 
\end{figure}

In Fig.~\ref{fig1}(a) 
we present ${{\cal N}}(\omega)$ for two different $J/t=0.3$ and
$J/t=0.6$ on a $N=18$ sites cluster for $V=0$. We observe that the
pseudo gap scales approximately as $2J$. The analysis at elevated
temperatures shows that the gap slowly fills up and disappears at $T
\sim J$. In Fig.~\ref{fig1}(b) results for different values $V$
are presented.  The gap remains robust also in the presence of the
$V$ term, which enhances ($V<0$) or suppresses ($V>0$) the
binding of hole pairs. It is thus evident that the effect of $V$ is
only of qualitative character and not relevant for the existence of
the pseudo gap. We therefore believe that this analysis suggests that
the origin of the pseudo gap is in short-range AFM spin correlations
rather than in the binding tendency of doped holes.

AFM spin correlations are correctly taken into account in the
SCBA. Therefore for the limiting case of low doping, $c_{\rm h}\to0$, 
and $T\to0$ we
approximate ${\cal N}^-(\omega)$ with 
\begin{equation}
{\cal N}^-(\omega)\propto \sum_{\bf k} {\rm Im} G_{\bf
k}(-\omega),
\end{equation}
where $G_{\bf k}(\omega)$ is the SCBA Green's function for adding one
hole (ARPES) to an AFM reference system (instead of adding one
hole to the state with one hole).
The corresponding DOS for {\it removing} a hole (IPES) from the
state with one hole, ${\cal N}^+(\omega)={2 \over N}
\sum_{\bf k} A_{\bf k}^+(\omega)$, can be calculated
accurately in the SCBA as follows.
First the spin averaged hole-like  spectral function,
\begin{equation}
A_{\bf k}^+(\omega)=-{1 \over 2\pi} {\rm Im}\sum_\sigma\langle \Psi_{{\bf k}_0}|
{ c_{{\bf k},\sigma}} {1 \over \omega-\tilde H}  c^\dagger_{{\bf k},\sigma}
|\Psi_{{\bf k}_0} \rangle,
\end{equation}
is expressed in terms of holon and magnon operators and the result
for ${\cal N}^+(\omega)$ emerges,
\begin{equation}
{\cal N^+(\omega)}= {1 \over N}\sum_i 
\langle \Psi_{{\bf k}_0}| h^\dagger_i[
{\delta(\omega-\tilde H)}+
a_i  {\delta(\omega-\tilde H)} a^\dagger_i] h_i 
|\Psi_{{\bf k}_0} \rangle.
\end{equation}
Here  $|\Psi_{{\bf k}_0} \rangle$
represents a weakly doped AFM, i.e., it is the ground state (GS)
wave function of a planar AFM with one hole and the GS wave vector
${\bf k}_0$.  The normalization (sum rule) 
of $A_{\bf k}^\pm(\omega)$ and ${\cal
N^\pm(\omega)}$ is discussed in detail in Ref.~\cite{jprev}. It should
be noted that the normalization of $A_{\bf k}^+(\omega)$ is not
trivial and is related to the EMD presented in the next section.

\begin{figure}[Htb]
\center{\epsfig{file=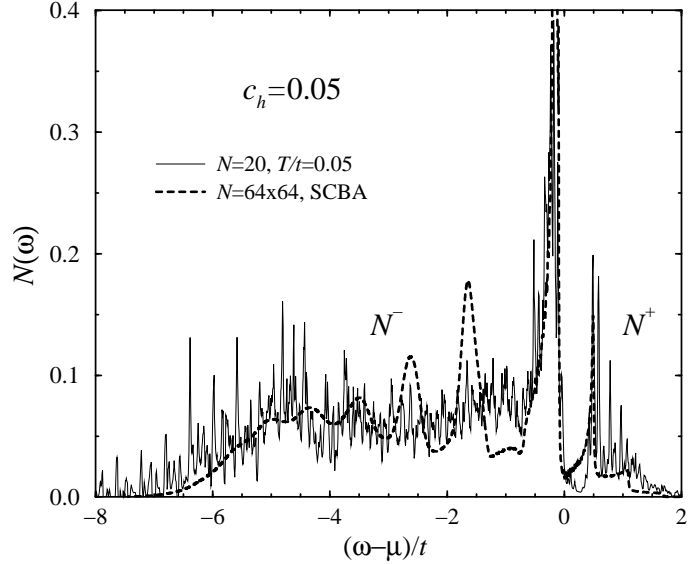,height=105mm,angle=-90}}
\caption{${\cal N}(\omega)$ for $N_h=1$ on $N=20$ sites, with
$J/t=0.3$, $V=0$, $T/t=0.05$ (full line). Dashed heavy line represents
the SCBA result on large lattice obtained as a sum of ${\cal
N}^-(\omega)$ and ${\cal N}^+(\omega)$. The SCBA result is obtained on
a $N=64\times 64$ cluster and for { undoped} reference system. Note the 
"string states" resonances, absent in the finite doping Green's
function. ${\cal N}^+(\omega)$ corresponds to IPES. Reference
hole concentration is $c_{\rm h}=1/N$. The SCBA result is normalized to
$c_{\rm h}=1/20$. Broadening of peaks is taken $\delta/t=0.01$.}\label{fig2}
\end{figure}

In Fig.~\ref{fig2} are shown spectra ${\cal N}(\omega)$ obtained with
the ED
on a $N=20$ sites cluster.   We compare these
spectra with the DOS within the SCBA.  The peaks in
${\cal N}^+(\omega)$ can well be explained with magnon structure of
single hole ground state, while peaks in ${\cal N}^-(\omega)$ are
string states known in the single hole case and are for the
present study of the pseudo gap and FS not relevant. 
As seen in Fig.~\ref{fig2} the total DOS obtained
with the SCBA and in particular ${\cal N}^+(\omega)$ remarkably accurately
resemble the ED DOS.

\section{Large or small Fermi surface?}

The electron momentum distribution function $n_{\bf k}=\langle
\Psi_{{\bf k}_0}|\sum_\sigma  c_{{\bf k},\sigma}^\dagger 
 c_{{\bf
k},\sigma} |\Psi_{{\bf k}_0} \rangle$ is the key quantity for
resolving the problem of the Fermi surface.
Numerically $n_{\bf k}$ can be determined by exact diagonalization
of the model in small clusters, where only a
restricted number of momenta ${\bf k}$ is allowed. The GS wave vector
due to finite size effects varies with $N$. Therefore we present here
results obtained with the method of twisted boundary conditions
\cite{zotos}, where $t_{jj'} \to t_{jj'} \exp{i {\theta}_{jj'}}$.
Since $n_{\bf k}\equiv n_{\bf k}({\bf k}_0,{\bf \theta})$ depends both
on ${\bf k}_0$ and ${\bf \theta}$ it follows from Peierls construction
that $n_{\bf k}({\bf k}_0,0)=n_{{\bf k}+{\bf k}_0}(0,{\bf k}_0)$ for
${\bf \theta}={\bf k}_0$. This allows us to study $n_{\bf k}$ for
arbitrary ${\bf k}$ and ${\bf k}_0$.  Furthermore, the finite size
effects of the results are suppressed if we fix ${\bf k}_0$ for {\it
all clusters} here studied to the symmetry point ${\bf
k}_0=(\frac\pi2,\frac\pi2)$.

\begin{figure}[Htb]
\center{\epsfig{file=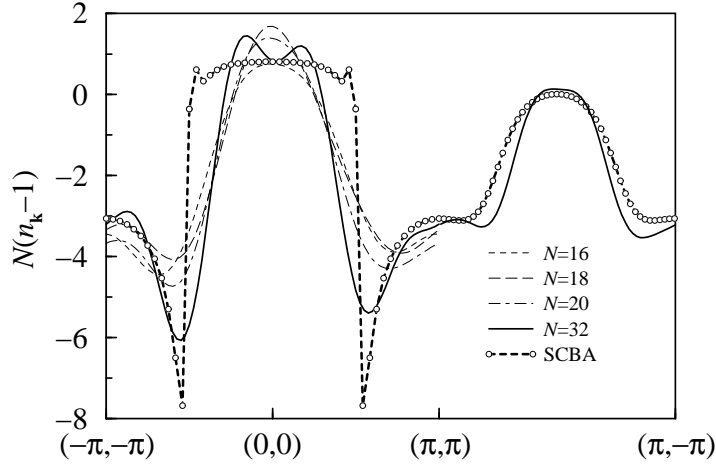,height=102mm,angle=-90}}
\caption{ $N (n_{\bf k}-1)$ obtained from ED 
for various systems $N=16, 18, 20, 32$ and $J/t=0.3$.
For the SCBA $N=64\times64$,
$\gamma=0.999$ and note that delta-function contributions 
at ${\bf k}=\pm {\bf   
k}_0$ are not shown. In plotting the curve for $N=32$ 
data from Ref.~\cite{chernyshev98} were used.}\label{fig3}
\end{figure}

In Fig.~\ref{fig3} we present for $J/t=0.3$ 
ED results for clusters with different $N$ and
$\gamma=1$. The EMD obeys the sum rule $\sum_{\bf k} n_{\bf k}=N-1$
and, for the allowed momenta, the constraint $N (n_{\bf
k}-1)\leq1$. We show here the quantity $N (n_{\bf k}-1)$, which for
different $N$ scales towards the same curve.  Results are presented
for some selected directions in the Brillouin zone (BZ) and should be
averaged over all four possible ground state momenta when discussed,
e.g., in connection with ARPES data.
  
Analytically we study the EMD again in the spinless fermion --
Schwinger boson representation.  The wave function Eq.~(\ref{psi})
corresponds to the projected space of the model Eq.~(\ref{tjv}) and
$n_{\bf k}=\langle \Psi_{{\bf k}_0}|
{\hat n_{\bf k}}|\Psi_{{\bf k}_0} \rangle$
with the
projected {\it electron} number operator ${\hat n_{\bf k}}=\sum_\sigma
{ c}_{{\bf k},\sigma}^\dagger  { c}_{{\bf k},\sigma}$.
Consistent with the SCBA approach, we decouple ${\hat n_{\bf k}}$
into hole and magnon operators,
\begin{equation}
{\hat n_{\bf k}}=\frac1N {\sum_{i j}}  h_i  h_j^\dagger
\biggl(\eta^+_{ij} \bigl[1+a_i^\dagger a_j (1-\delta_{ij})
\bigr]
+ \eta^-_{ij}(a_i^\dagger + a_j)\biggr),\label{nk}
\end{equation}
where $\eta^\pm_{ij}=e^{-i{\bf k} \cdot
({\bf R}_{i}-{\bf R}_{j})}(1\pm e^{-i{\bf Q} \cdot ({\bf R}_{i}-{\bf
R}_{j})})/2$ with ${\bf Q}=(\pi,\pi)$.  Local $a^\dagger_i$ are
further expressed with proper magnon operators $\alpha^\dagger_{\bf
q}$. In general the expectation value $n_{\bf k}$ for a single hole
has the following structure \cite{ramsak10_ram2000}
\begin{equation}
n_{\bf k}=1-\frac12 Z_{{\bf k}_0}(\delta_{{\bf k} {\bf k}_0}+
\delta_{{\bf k} {\bf k}_0+{\bf Q}})+{1 \over N}
\delta n_{\bf k}.\label{dnk}
\end{equation}
Here the second term proportional to $\delta$-functions corresponds
to hole pockets. Note that $\delta n_{\bf
k}$, for the case of a single hole fulfills the sum
rule $\frac1N\sum_{\bf k} \delta n_{\bf k}=Z_{{\bf k}_0}-1$ and
$\delta n_{\bf k}\le1$. The introduction of $\delta n_{\bf k}$ is
convenient as it allows the comparison of results obtained with
different methods and on clusters of different size $N$.

In Fig.~\ref{fig3} we also present the SCBA result. We have also
checked the convergence of $\delta n_{\bf k}$ with the number of
magnon lines, $n$.  For $J/t >0.3$ we find for all ${\bf k}$ that the
contribution of terms $n>3$ amounts to less than few percent. This is
in agreement with the convergence of the norm of the wave function,
which is even faster \cite{ramsak93}.
 
\begin{figure}[Htb]                     
\center{\epsfig{file=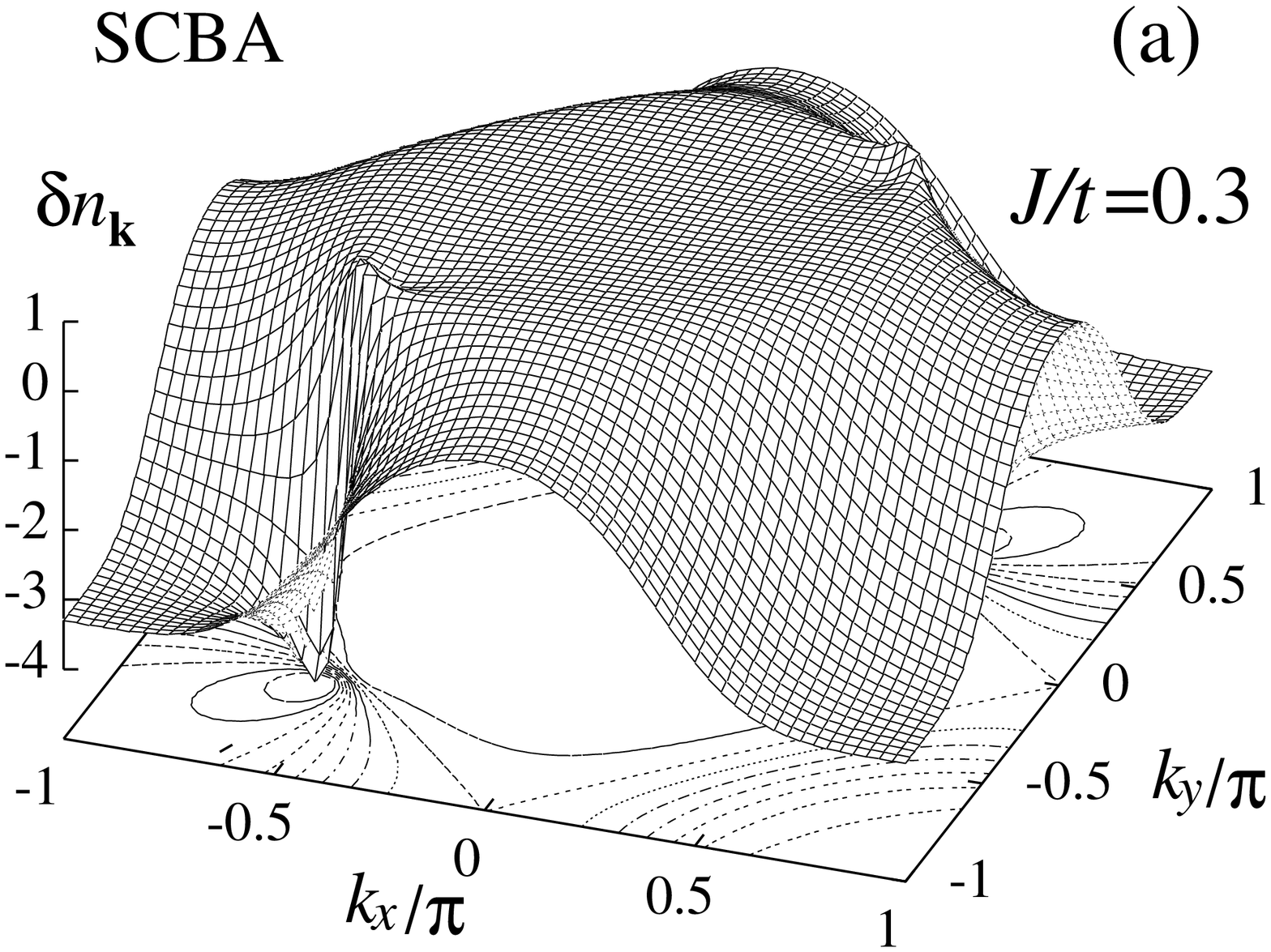,height=59mm,angle=-90}   
\epsfig{file=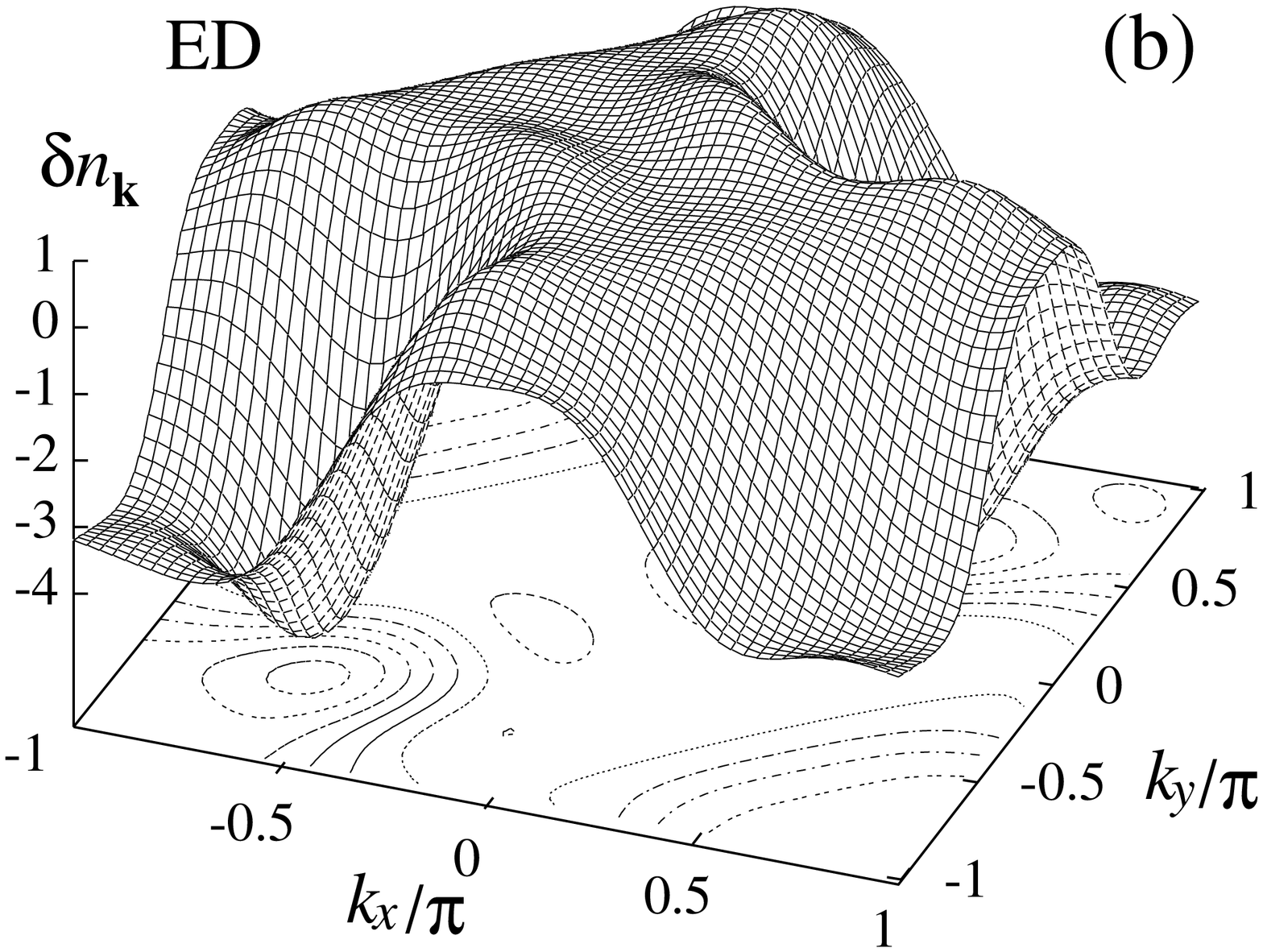,height=59mm,angle=-90}}         
\caption{(a) SCBA result for $N=64\times64$ and $J/t=0.3$, $\gamma=0.99$. (b)
Exact diagonalization result for $N=32$ sites as in Fig.~\ref{fig3}.}
\label{fig4}
\end{figure}

In Fig.~\ref{fig4}(a) we present this $\delta
n_{\bf k}$ for the whole BZ. The
important ingredient of the SCBA is the gapless magnons
with linear dispersion and a more complex ground state of the planar
AFM.  $G_{\bf k}(\omega)$ and $\epsilon_{\bf k}$ are strongly ${\bf
k}$-dependent.  As a consequence $n_{\bf k}$ in general
depends both on ${\bf k}$ and ${\bf k}_0$. The ground
state is for the $t$-$J$ model fourfold degenerate and we again choose ${\bf
k}_0=(\frac\pi2,\frac\pi2)$. In Fig.~\ref{fig4}(b) is presented also
the result of exact
diagonalization of a $N=32$ sites cluster \cite{chernyshev98}, but 
generalized to the whole BZ.

The main conclusion regarding the EMD is the coexistence of two
apparently contradicting Fermi-surface scenarios in EMD of a single
hole in an AFM. (i) On one hand, the $\delta$-function contributions
in Eq.~(\ref{dnk}) seem to indicate that at finite doping a delta-function
might develop into small Fermi surface, i.e., a hole pocket, provided
that AFM long range order persists. (ii) A novel feature is that also
$\delta n_{\bf k}$ is singular in a particular way, i.e., it shows a
discontinuity at ${\bf k}={\bf k}_0$ with a strong asymmetry with
respect to ${\bf k}_0$. It is therefore more consistent with
infinitesimally short arc (point) of an emerging large FS. For finite
doping the discontinuity could possibly extend into such a finite arc
(not closed) FS. Note that as long-range AFM order is destroyed by
doping, hole pocket contributions should disappear while the
singularity in $\delta n_{\bf k}$ could persist.  The results of the
two methods, the SCBA and the ED
agree quantitatively at all points in the BZ. However,
the SCBA result is {\it symmetric} around $\Gamma$ point in the
direction ${\bf k}\parallel {\bf k}_0$, while small system results
show a weak asymmetry for ${\bf k}=\pm {\bf k}_0$, respectively. From
our analysis of the SCBA results for $N\to \infty$ and long range AFM
spin background it follows that in the thermodynamic
limit $c_{\rm h}\to0$ $n_{\bf k}$ is symmetric.  The asymmetry is in
Ref.~\cite{chernyshev98} attributed to the opening of the gap in the
magnon spectrum at ${\bf q}\sim{\bf Q}$ in finite systems. Within the
SCBA the asymmetry also appears if the EMD is evaluated with ${\bf
k}_0$ {\it displaced} from $(\frac\pi2,\frac\pi2)$ by a small amount
$\delta{\bf k}_0$ (not shown here).

\section{Summary and analytical results}

Full numerical results are captured with a simple analytical
expansion which gives more insight into the structure of $A_{\bf
k}^+(\omega)$ and $\delta n_{\bf k}$. We simplify the wave function,
Eq.~(\ref{psi}), by keeping only the one-magnon contributions and take
$J/t \gg 1$ and the leading order contributions are then
\begin{eqnarray}A_{\bf k}^+(\omega) &\sim&
A_{\bf k}^{(1)}(\omega)=
[ Z_{{\bf k}_0}\delta_{{\bf q},0}
+{1 \over N}
(1-\delta n_{\bf k}^{(1)})]\delta(\omega-\omega_{\bf q}),
\nonumber\\
\delta n_{\bf k}^{(1)}\!\!&=&\!-Z_{{\bf k}_0}M_{{\bf k}_0{\bf q}}
G_{{\bf k}_0}(\epsilon_{{\bf k}_0}\!\!-\!\omega_{\bf q})\bigl[2 u_{\bf q}\!+\!
M_{{\bf k}_0{\bf q}}
G_{{\bf k}_0}(\epsilon_{{\bf k}_0}\!\!-\!\omega_{\bf q})\bigr] \nonumber\\
&\sim& -8  Z_{{\bf k}_0}^2 J
{{\bf q}\cdot {\bf v} \over \omega^2_{\bf q}}
(1+ Z_{{\bf k}_0} {{\bf q}\cdot {\bf v}  \over
\omega_{\bf q}  }), \qquad {q\to0}, \label{dn1}
\end{eqnarray}
\noindent
with ${\bf q}={\bf k}-{\bf k}_0$ (or ${\bf k}-{\bf k}_0-{\bf Q}$) and
${\bf v}=t(\sin k_{0x},\sin k_{0y})$. 

A surprising observation is that the EMD exhibits for momenta ${\bf
k}\sim {\bf k}_0, {\bf k}_0+{\bf Q}$ a discontinuity $\sim Z_{{\bf
k}_0} N^{1/2}$ and $\delta n_{\bf k}^{(1)} \propto -(1+{\rm sign}\,
q_x)/q_x$. These discontinuities are consistent with ED results. One
can interpret this result as an indication of an emerging {\it large}
Fermi surface at ${\bf k}\sim \pm {\bf k}_0$.  The discontinuity
appears only as {\it points} $\pm {\bf k}_0$, not {\it lines} in the
BZ. Note, however, that this result is obtained in the extreme low
doping limit, i.e., $c_{\rm h}=1/N$ and it is not straightforward to
generalize it to the finite doping regime.
  
\begin{figure}[Htb]        
\center{\epsfig{file=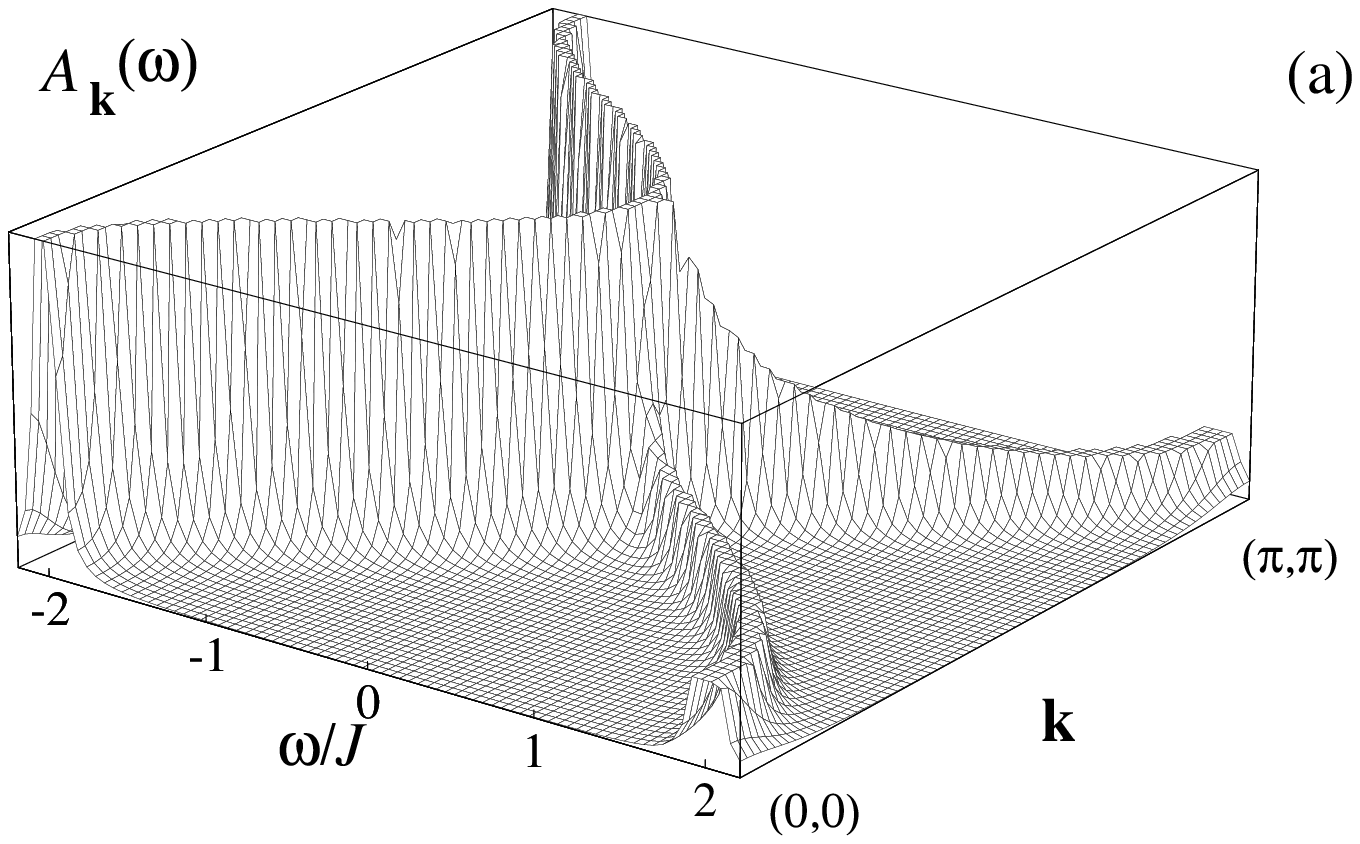,height=47mm,angle=0}   }   
\vskip -1.5 cm \hskip 0 cm                 
\center{\epsfig{file=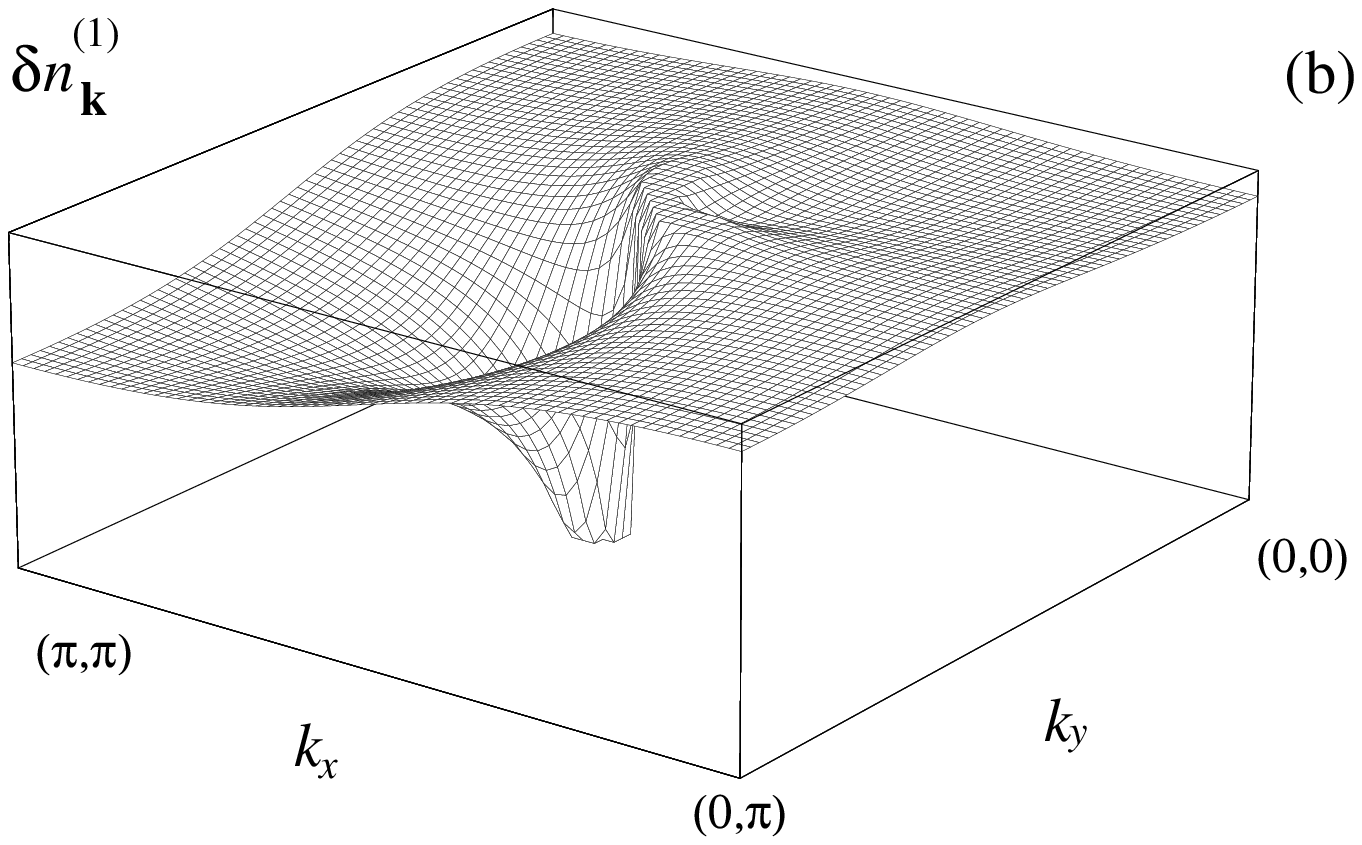,height=47mm,angle=0}   }
\caption{The tendency of formation of the large FS as seen in
(a) $A_{\bf k}(\omega)$  and
(b) $\delta n_{\bf k}^{(1)}$ for $Z_{{\bf k}_0}t/J\sim1$. 
The figures are in arbitrary units and
clipped for convenience.
}\label{fig5}
\end{figure}

A direct reflection of the anomaly in $\delta n_{\bf k}$ is also the
structure of $A_{\bf k}(\omega)$ as presented in Fig.~\ref{fig5}(a),
where $A_{\bf k}^+(\omega)$ exhibits a typical asymmetry in momentum
dependence. This shows the instability towards the large FS.
The electron part is in this figure approximated with $A_{\bf
k}^-(\omega) \sim Z_{{\bf k}_0} \delta(\epsilon_{\bf k}-\epsilon_{{\bf
k}_0}-\omega)$. In Fig.~\ref{fig5}(b) $\delta n_{\bf k}^{(1)}$
is presented and the anomaly discussed above is clearly seen.

We conclude by stressing that the origin of the pseudo
gap found in cuprates seems to be in the short range spin correlations
of the reference AFM system, as well as in the strong asymmetry
between the hole-like and electron-like spectra in underdoped
systems. From the present SCBA analysis it is clear that the gap size
is a natural consequence of magnons with dispersion $\sim 2J$, and the
first peak in ${\cal N}^+(\omega)$ thus corresponds to the "van
Hove" high density of magnon states.
In addition in
making contact with ARPES experiments we should note that ARPES
measures the imaginary part of the electron Green's function.  We must
note that using these experiments in underdoped cuprates $n_{\bf k}$
can be only qualitatively discussed since the latter is extracted only
from rather restricted frequency window below the chemical potential.
Nevertheless our results are not consistent with a small hole pocket
FS (at least only a part of presumable closed FS is visible), but
rather with partially developed arcs resulting in FS which is just a
set of disconnected segments at low temperature collapsing to the
point \cite{jcc,norman98}. The SCBA results for singular $\delta n_{\bf
k}$ seem to allow for such a scenario. It should also be stressed that
the SCBA approach is based on the AFM long-range order, still we do
not expect that finite but longer-range AFM correlations would
entirely change our conclusions.

\vskip -.5 cm \hskip 0 cm                 
\center{\epsfig{file=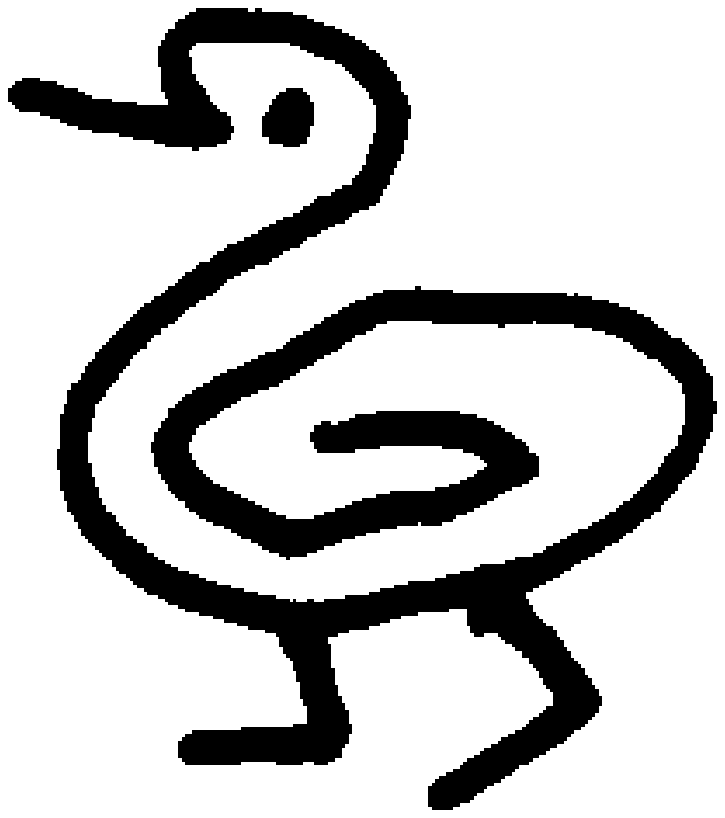,height=5mm,angle=0}   }   

\begin{chapthebibliography}{99}        
\bibitem{jcc} {J.C. Campuzano and M. Randeria, 
this volume, p. 11.}
\bibitem{shen95} {Z.-X. Shen and D.S. Dessau, Phys. Rep. {\bf 253}, 1
(1995); B.O. Wells {\it et al.}, Phys. Rev. Lett. {\bf
74}, 964 (1995); D.S. Marshall {\it et al.},
Phys. Rev. Lett. {\bf 76}, 4841 (1996).}
\bibitem{norman98} {M.R. Norman {\it et al.}, Nature {\bf 392}, 157 (1998).}
\bibitem{imada} {For a review see, e.g., M. Imada, A. Fujimori, and Y.
Tokura, Rev. Mod. Phys. {\bf 70}, 1039 (1998).}
\bibitem{dago} {For a review see, e.g., E. Dagotto,
Rev. Mod. Phys. {\bf 66}, 763 (1994).} 
\bibitem{stephan91} {W. Stephan and P. Horsch, Phys. Rev. Lett. {\bf
66}, 2258 (1991).}
\bibitem{eder98} {R. Eder and Y. Ohta, Phys. Rev. B {\bf 57}, R5590 (1998).}
\bibitem{chernyshev98} {A.L. Chernyshev, P.W. Leung and
R.J. Gooding, Phys. Rev. B {\bf 58}, 13594 (1998).}
\bibitem{eder91} {R. Eder, Phys. Rev. B, {\bf 44}, R12609 (1991).}
\bibitem{lee96} {X.-G. Wen and P.A. Lee, Phys. Rev. Lett. {\bf 76}, 503
(1996).}
\bibitem{putikka98} {W.O. Putikka, M.U. Luchini, and R.R.P. Singh,
J. Phys. Chem.  Solids {\bf 59}, 1858 (1998); Phys. Rev. Lett. {\bf
81}, 2966 (1998).}   
\bibitem{furukawa98} {N. Furukawa, T.M. Rice, and M. Salmhofer,
Phys. Rev. Lett., {\bf 81}, 3195 (1998).}
\bibitem{jprev} {For a review see J. Jakli\v c and P.  Prelov\v sek,
Adv. Phys.  {\bf 49}, 1 (2000).}
\bibitem{jpspec} {P. Prelov\v sek, J. Jakli\v c, and K. Bedell,
Phys. Rev. B {\bf 60}, 40 (1999).}
\bibitem{sigmac} {P. Prelov\v sek, A. Ram\v sak, and I. Sega,
Phys. Rev. Lett. {\bf 81}, 3745 (1998).}
\bibitem{schmitt88} {S. Schmitt-Rink, C.M. Varma, and A.E. Ruckenstein,
Phys. Rev. Lett. {\bf 60}, 2793  (1988).}
\bibitem{ramsak90} {A. Ram\v sak and P. Prelov\v sek,
Phys. Rev. B {\bf 42}, 10415 (1990).}
\bibitem{martinez91} {G. Mart\'{\i}nez and P. Horsch, Phys. Rev.  B {\bf 44},
317 (1991). }
\bibitem{reiter94} {G.F. Reiter, Phys. Rev. B {\bf 49}, 1536 (1994).}
\bibitem{ramsak93} {A. Ram\v sak and P. Horsch, Phys. Rev. B {\bf 48},
10559 (1993); {\it ibid.} {\bf 57}, 4308 (1998).}
\bibitem{bala95} {J. Ba\l a, A. M. Ole\' s, and J. Zaanen,
Phys. Rev. B {\bf 52}, 4597 (1995).}
\bibitem{zotos} {X. Zotos, P. Prelov\v sek and I. Sega, Phys. Rev. B
{\bf 42}, 8445 (1990).}
\bibitem{ramsak10_ram2000} {A. Ram\v sak, P. Prelov\v sek, and
I. Sega, Phys. Rev. B {\bf 61},
4389 (2000).}
\end{chapthebibliography}
\end{document}